# Mass & Light in Galaxy Clusters: The Case of Abell 370

Marceau Limousin[1], Benjamin Beauchesne[2,3], Anna Niemiec[4], Jose M. Diego[5], Mathilde Jauzac[6,7,8,9], Anton Koekemoer[10], Keren Sharon[11], Ana Acebron[3], David Lagattuta[4,5], Guillaume Mahler[12,4,5], Lilia L. R. Williams[13], Johan Richard[14], Eric Jullo[1], Lukas J. Furtak[15], Andreas L. Faisst[16], Brenda L. Frye[17], Pascale Hibon[10], Priyamvada Natarajan[18,19], Michael Rich[20] *

[1] Aix Marseille Univ, CNRS, CNES, LAM, Marseille, France
[2] Institute of Physics, Laboratory of Astrophysics, Ecole Polytechnique Fédérale de Lausanne (EPFL), Observatoire de Sauverny, 1290 Versoix, Switzerland
[3] ESO, Alonso de Córdova 3107, Vitacura, Santiago, Chile
[4] LPNHE, CNRS/IN2P3, Sorbonne Université, Université Paris Cité, Laboratoire de Physique Nucléaire et de Hautes Energies, 75005 Paris, France
[5] Instituto de Fisica de Cantabria (CSIC-UC), Avda. Los Castros s/n, E-39005 Santander, Spain
[6] Centre for Extragalactic Astronomy, Department of Physics, Durham University, Durham DH1 3LE, UK
[7] Institute for Computational Cosmology, Durham University, South Road, Durham DH1 3LE, UK
[8] Astrophysics Research Centre, University of KwaZulu-Natal, Westville Campus, Durban 4041, South Africa
[9] School of Mathematics, Statistics & Computer Science, University of KwaZulu-Natal, Westville Campus, Durban 4041, South Africa
[10] Space Telescope Science Institute, 3700 San Martin Drive, Baltimore, MD 21218, USA
[11] Department of Astronomy, University of Michigan, 1085 S. University Ave, Ann Arbor, MI 48109, USA
[12] STAR Institute, Quartier Agora - Allée du six Août, 19c B-4000 Liège, Belgium
[13] School of Physics and Astronomy, University of Minnesota, Minneapolis, MN, 55455, USA
[14] Université Lyon 1, CNRS, Centre de Recherche Astrophysique de Lyon (CRAL), Saint-Genis-Laval, France
[15] Physics Department, Ben-Gurion University of the Negev, P.O. Box 653, Be'er-Sheva 84105, Israel
[16] Caltech/IPAC, 1200 E. California Blvd. Pasadena, CA 91125, USA
[17] Department of Astronomy/Steward Observatory, University of Arizona, 933 N. Cherry Avenue, Tucson, AZ 85721, USA
[18] Department of Astronomy, Yale University, P.O. Box 208101, New Haven, CT 06520, USA
[19] Department of Physics, Yale University, P.O. Box 208120, New Haven, CT 06520, USA
[20] Department of Physics and Astronomy, University of California, Los Angeles, CA 90095-1547, USA

Preprint online version: October 24, 2024

**ABSTRACT**

In the cold dark matter paradigm, the association between the hypothetic dark matter and its stellar counterpart is expected. However, parametric strong lensing studies of galaxy clusters often display "misleading features": group/cluster scale dark matter components without any stellar counterpart, offsets between both components larger than what might be allowed by neither CDM nor self interacting DM models, or significant unexplained external shear components. This is the case in the galaxy cluster Abell 370, which mass distribution has been the subject of several studies motivated by a wealth of data. Using strong lensing techniques, it has been described parametrically by a four dark matter clumps model and galaxy scale perturbers, as well as a significant external shear component, which physical origin remains a challenge. The dark matter distribution features a mass clump with no stellar counterpart and a significant offset between one of the dark matter clumps and its associated stellar counterpart. In this paper, based on BUFFALO data, we begin by revisiting this mass model. Sampling this complex parameter space with MCMC techniques, we find a four dark matter clumps solution which does not require any external shear and provides a slightly better RMS compared to previous models ($0.7''$ compared to $0.9''$). Investigating further this new solution, in particular playing with the parameters leading the MCMC sampler, we present a class of models which can accurately reproduce the strong lensing data, but whose parameters for the dark matter component are poorly constrained, limiting any insights on its properties. We then develop a model where each large scale dark matter component *must* be associated with a stellar counterpart. This three dark matter clumps model is unable to reproduce the observational constraints with an RMS smaller than $2.3''$, and the parameters describing this dark matter component remain poorly constrained. Examining the *total* projected mass maps, we find a good agreement between the total mass and the stellar distribution, both being, to first order, bimodal. We interpret the "misleading features" of the four dark matter clumps mass model and the failure of the three dark matter clumps mass model as being symptomatic of the lack of realism of a parametric description of the dark matter distribution in such a complex merging cluster, and encourage caution and criticism on the outputs of parametric strong lensing modelling.

We briefly discuss the implications of our results for using Abell 370 as a gravitational telescope. Using the class of models reproducing the strong lensing data, we compute the magnifications for background Ly$\alpha$ emitters, and present the critical curves obtained for the redshift of the "Dragon" arc whose recent JWST observations prompted interest. Finally, in light of our results, we discuss the strategy of choosing merging (multimodal) clusters as gravitational telescopes compared to simple (unimodal) clusters.

## 1. Mass & Light in Galaxy Clusters

Both observations and numerical simulations do support the association between mass and light at the galaxy cluster scale. Here, we refer to mass as being the largely dominating dark matter (DM) component whose existence remains to be confirmed. We refer to light as being the associated stellar component, in most cases in the form of the brightest cluster galaxy (BCG). Observationally, no cluster scale DM clump without any associated light concentration has been reliably detected so far. Besides, in hydrodynamical simulations, stars do form in the potentiel well of DM halos. This results into the hierarchical formation of a bright galaxy found at the centre of the underlying DM halo.

If DM is collisionless as proposed in the Cold Dark Matter (CDM) scenario, the association between mass and light should be perfect, *i.e.* the offset between the peaks of each component should be equal to 0 (Roche et al. 2024). If DM is self interacting, such an offset is possible and should be at most of the order of a few dozen of kpc, according to simulations (Kim et al. 2017; Robertson et al. 2017; Tulin & Yu 2018; Fischer et al. 2021; Adhikari et al. 2022).

Observing offsets between DM and light would therefore be an interesting probe of the DM properties, in particular its possible self interaction. Since DM is more likely to interact in colliding clusters than in relaxed clusters, offsets are more likely to be found in merging clusters. DM should experience a drag force (similar to the one experienced by the X-ray emitting gas but with a much smaller amplitude) which does not affect the stellar component, and thus will lead to a separation between the stellar and the DM components. This configuration has been recently studied numerically, quantifying this offset for different values of the self interaction cross section of DM particles (Sirks et al. 2024). For a vanishing cross section (corresponding to the CDM case), a spatial offset consistent with 0 is retrieved. For cross sections equal to 0.1 and $1\,\mathrm{cm}^2/\mathrm{g}$, a mean offset of 5 and 12 kpc is respectively derived by these authors. This is the same order of magnitude of what is found in recent "El Gordo" like simulations by Valdarnini (2024): 28 kpc with a cross section equal to $1\,\mathrm{cm}^2/\mathrm{g}$.

Observationally, no such offset has been reliably detected yet. In practice, we need to be able to constrain the position of the peak of the DM component with sufficient accuracy, *i.e.* smaller than the typical offset we aim to detect. Gravitational lensing is a useful method in this respect. Weak gravitational lensing typical positional uncertainty is still of order of $\sim 10''$ (which translates into $\sim 45\,\mathrm{kpc}$ for a cluster at redshift 0.3) using the current available data (Kim et al. 2017, 2021). Strong lensing (SL) is much more promising in this respect, with current typical uncertainties being at the sub-arcsec level. However, multiple image misidentification can complicate the situation (see the works by Massey et al. 2015, 2018, in Abell 3827), as well as the assumed lens properties (see Lin et al. 2023, in Abell 3827).

In the case of the merging Bullet Cluster, where DM clump positions are well constrained by SL, no such offset has been detected, providing an upper limit on the self interacting cross section of DM equal to $1.25\,\mathrm{cm}^2/\mathrm{g}$ (Randall et al. 2008). The non detection of such an offset in a sample of 72 colliding clusters by Harvey et al. (2015) led to an upper limit equal to $0.47\,\mathrm{cm}^2/\mathrm{g}$.

Beyond the offset between a DM clump and its associated stellar component is the notion of "dark clump", *i.e.* group/cluster scale DM clump with *no* associated luminous counterpart. As mentioned earlier, a dark clump has never been reliably detected so far but such features are seen in parametric SL mass modelling of clusters, both unimodal and multimodal. The physical interpretation of such dark clumps can be misleading and limits the advantages of parametric SL modelling, that is to compare the description of the mass clumps inferred through SL analysis with the theoretical expectations. We therefore consider it worthwhile to look more closely into the details at SL mass models where such "misleading features" are reported. In Limousin et al. (2022, L 22 hereafter), we revisited three parametric mass models where previous (sometimes independent) analysis displayed dark clumps. The three clusters in that study are AS 1063, MACS J0416 and MACS J1206. We presented competitive mass models which did not require the inclusion of dark clumps. We proposed a working hypothesis where each DM clump should coincide with the associated stellar component within a few dozen kpc, the typical offset allowed by SIDM models.

Following this approach, we revisit here the mass model of Abell 370, for which parametric studies suggest the presence of DM not associated with light. Our aim is to see if we can propose a mass model where the DM component is traced by light.

We begin by reviewing the parametric mass modellings performed on Abell 370 in Section 2, revisiting the current four DM clumps mass model in Section 3 and 4. We then propose a three DM clumps mass model where mass is traced by light in Section 5. We discuss our results and conclude in Section 6, and briefly present the implication of our findings for high redshift studies in Section 7.

## 2. Mass & Light in Abell 370

Abell 370, a rich galaxy cluster located at $z = 0.375$, is of historical relevance when it comes to strong lensing (SL), since this is the cluster in which the first giant gravitational arc was discovered (Lynds & Petrosian 1986; Soucail et al. 1987; Lynds & Petrosian 1989). A first look at an optical image of the core of Abell 370 reveals a complex light distribution, suggesting a complex underlying mass distribution (see Fig. 1, as well as the smoothed light map presented in Lagattuta et al. 2017) resulting from an ongoing merger (Molnar et al. 2020).

The light distribution is dominated by the light associated with two dominant bright galaxies, that we will refer to as the northern and the southern BCG (BCG-N and BCG-S hereafter). We also observe a "crown" of bright galaxies surrounding BCG-N, a light concentration in the east/north-east (East component hereafter), as well as another one further north.

*Parametric* mass models based on the observation of multiple images have been proposed in the past. We review here the ones that are the most relevant for this work, in particular on the description of the smooth dark matter distribution. As is usually done, on top of this smooth component, cluster members are added, using scaling laws to re-late their total mass to their luminosity (see, *e.g.* Limousin et al. 2007).

- ACS multiband images were obtained as part of the Early Release Observation that followed the Hubble Service Mission #4. Based on these data, Richard et al. (2010), reported 9 multiply imaged systems. Describing the cluster with a bimodal mass distribution, *i.e.* two DM clumps associated with each BCG, they are able to reproduce the multiple images with an RMS equal to $1.76''$. The northern DM clump is found to be located at $\sim 10''$ from BCG-N.
- Richard et al. (2014), analysing new WFC3 data, reported 12 systems. The description of the mass distribution remains bimodal. The SL optimisation is performed in the source plane. This mass model was constructed in light of the forthcoming Hubble Frontier Fields (HFF) data.
- The wealth of new constraints revealed by the depth of the HFF data motivated several groups to propose new mass models. In particular, Lagattuta et al. (2017) proposed a mass model which, compared to pre-HFF data, includes two extra DM haloes, resulting in a four DM clumps mass model. One of them is associated with BCG-S, one with BCG-N and one with the eastern light component. The fourth one is not associated with any light concentration and ends up between BCG-S and BCG-N, called "the bridge". As in Richard et al. (2010), the DM clump associated with BCG-N ends up at $\sim 10''$ from BCG-N. They report an RMS equal to $0.94''$.
- Additional data obtained with MUSE provided more constraints, and the number of spectroscopically confirmed images doubled. The mass model was revisited by Lagattuta et al. (2019, L19 hereafter). It still consists of four large scale DM clumps. They also optimise individually a few cluster members: BCG-S and BCG-N, as well as 4 cluster members (green circles in Fig. 1) located close to multiple images and that are important to reproduce their observed positions accurately. The authors also introduced a strong external shear component (an amplitude of order 0.1) in order to lower the RMS. They tried to find a physical origin to this external shear component but did not find any. They also tried to add another DM clump associated with the light distribution located further north, but its velocity dispersion was converging towards 0, suggesting that it is not required.
- Niemiec et al. (2023, N23 hereafter,) used BUFFALO (Steinhardt et al. 2020) data in order to pursue a combined strong and weak lensing analysis of Abell 370. On the SL side, the main difference between N23 and L19 are small changes in the multiple image and cluster member catalogs. N23 considers the "gold" BUFFALO catalog of multiple images (the best, most confident, multiply imaged systems, with very low positional uncertainty and definite spectroscopic detection), as well as the BUFFALO cluster members catalog. The description of the mass distribution and the number of optimised components is the same as in L19. They report an RMS equal to $0.90''$. Using WL, they detect substructures in the outskirts. The inclusion of these substructures in the outskirts does not help to explain the external shear component which remains present in their model. Using extensive spectroscopic data covering the BUFFALO field of view, Lagattuta et al. (2022) did not find either any clear origin for this shear component, that we do question in this work.
- Recently, Gledhill et al. (2024) reported an updated model using JWST data. They report no new spectroscopically confirmed multiply imaged systems and find that the overall shape of the critical curve is similar to previous models. Their parametric model features *five* large scale DM haloes whose positions are not given in the paper.
- More recently, Li et al. (2024) proposed a new mass model in order to study the transient events detected in the giant arc thanks to the new JWST data. They first construct a "preliminary lens model", using the same set of multiple images as L19. They report that a mass model composed of two (NFW) large scale DM haloes (plus individual galaxies) is unable to reproduce the multiple images. These observational constraints are however well reproduced (RMS equal to $0.88''$) by a three clumps mass model, and they report that including a forth one provided no improvement, contrary to what is claimed by L19 and N23. Their model includes an external shear component with a much smaller amplitude (0.03) compared to what is required by L19 and N23. Looking at the *positions* of these three NFW DM mass clumps shows that *none* of them is actually associated with a luminous counterpart. However, since they add the cluster members in their model, the *total* mass distribution might follow the light distribution, but this information is not given in the paper, which concentrates on the giant arc. Their second NFW mass "clump" has a "concentration parameter" equal to *0.28*, which *never* happens in *any* numerical simulation. At the same time, they also individually optimize the mass associated with both BCGs. If the halo associated with BCG-S has a position coincident with the light peak of the latter, the halo associated with BCG-N has a position which is at $12''$ from the light peak of BCG-N, coincident with the position of their third DM halo. They then add multiple images located in the giant arc to improve their model in this area.

To summarize, if parametric mass modelling is able to reproduce the wealth of spectroscopically confirmed multiple images with a good accuracy (RMS<$1''$), *all* these models do not fully associate DM with light.

## 3. Revisiting the four DM clumps mass model

### 3.1. Starting Point

The starting point of this study is the SL model described in N23 and illustrated in Fig. 1. It reproduces the observational constraints with an RMS equal to $0.90''$, which is comparable with that of L19 ($0.78''$). It has been optimised in the image plane. In all this work, we use the same sets of multiply imaged systems and cluster members as in N23. Besides, all models are optimised in the image plane, using the LENSTOOL software (Jullo et al. 2007).

### 3.2. Rerunning the N23 model

We first rerun twice exactly the same run as the one presented in N23, *i.e.* using the very same LENSTOOL input files (parameter file, cluster members catalog and multiple image file). We reach RMS equal to $0.76''$ and $0.80''$. The values of the individual parameters describing the DM distribution differ. We will come back to this issue later in this paper.

This model displays a significant external shear, with an amplitude of order 0.1, comparable to that of N23. To put in context, this shear amplitude is what would be experienced at $\sim 150''$ from the centre of galaxy cluster Abell 1689 (Limousin et al. 2007). If we remove this external shear component from the model, we obtain an RMS equal to $1.01''$. We will refer to this last model (without external shear) as *model (i)* in the following.

### 3.3. Priors on the galaxy scale component

The degeneracies between the smooth DM component and the galaxy scale component in SL studies are well known. We refer the reader to Limousin et al. (2016) and L22 for a discussion. By measuring the mass associated with the cluster members, one can break this degeneracy, as proposed by Bergamini et al. (2019, B19 hereafter) and more recently by Beauchesne et al. (2024).

Indeed, using MUSE spectroscopy, B19 were able to provide priors on the scaling laws used in the SL modelling that relates the mass associated with a given cluster member with its luminosity. A similar study is not yet available for Abell 370, and we extrapolate the results by B19 on Abell 370. From the sample of three clusters studied by B19, we use the results derived on MACS 0416. Located at $z = 0.39$, it is the cluster of the B19 sample which is the closest in redshift with Abell 370 ($z = 0.37$). Besides, amongst the B19 sample, it is the cluster which dynamical state is the closest to Abell 370, *i.e.* presenting a multimodal light and mass distribution.

B19 propose a Gaussian prior on the velocity dispersion of a mag=17.05 (F160W) magnitude equal to $\sigma = 248 \pm 28$ km s$^{-1}$. We do impose a Gaussian prior equal to $248 \pm 50$ km s$^{-1}$, broadening the error bar in order to account for the differences between each cluster galaxy populations. To be consistent with B19, we modify the galaxy catalog by N23 and assign a F160W magnitude to each member galaxy. In the following, we refer to this prior on the galaxy scale component as the *"spectroscopic prior"*. Doing so, we obtain an RMS equal to $0.86''$, slightly larger but similar to the run with broad priors on the galaxy scale perturbers. A significant external shear component is favored, equal to 0.1, similar to former models discussed above. Values of the individual parameters describing the DM distribution differ. If imposing this spectroscopic prior degrades a bit the RMS, we consider worth including it since it is physically and observationally motivated.

Note that, following L19 and N23, for all models presented in this paper, we also optimise individually 4 cluster members as well as BCG-N and BCG-S (green circles on Fig. 1).

### 3.4. Removing the external shear component

Keeping the spectroscopic prior, we then remove the external shear component in the modelling. We obtain an RMS of $0.70''$, finding that the inclusion of an external shear component is not necessary. We will refer to this model as *model (ii)* in the following. Once again, values of the individual parameters describing the DM distribution differ.

### 3.5. Comparing models

We remind that the reference model is obtained by rerunning the N23 files, and has an RMS equal to $0.76''$ and an external shear of $\sim 0.1$. Then we derive the following models:
- *model (i)*: the reference model without external shear, with an RMS equal to $1.01''$.
- *model (ii)*: the reference model without external shear including the spectroscopic priors, with an RMS equal to $0.70''$. It constitutes the best model we have been able to derive in this paper, in terms of RMS. Moreover, it does not require the inclusion of a significant external shear component that cannot be physically explained, and it incorporates a physically motivated prior on the galaxy scale component.

It is interesting to note that *model (ii)* is contained in *model (i)*. Indeed, in *model (i)*, the priors on the galaxy scale component are large enough to contain the spectroscopic priors incorporated in *model (ii)*. Therefore, in *model (i)*, the sampler could have converged towards *model (ii)*, but it did not, although the fit is better. Actually, the priors in L19 and N23 mass models were large enough to find *model (ii)*. This is puzzling and interesting, and motivates further investigations in order to understand the situation better. This is the subject of the next Section.

## 4. Further investigating the four DM clumps mass model

Motivated by these findings, we decide to further investigate the four DM clumps mass model. We recall, here, that for the different models investigated above, the values of the individual parameters describing the four DM mass clumps differ. These are hints for checking the convergence of the Monte Carlo Markov Chain (MCMC) sampler.

Dark matter mass clumps are described using a dual Pseudo Isothermal Elliptical Mass Distribution (dPIE profile). We refer the reader to Limousin et al. (2005) and Elíasdóttir et al. (2007) for a description of this mass profile. Here we only give a brief overview. The geometrical parameters are the position, ellipticity and position angle. Then it is parametrised by a fiducial velocity dispersion, $\sigma$, a core radius, $r_{\rm core}$, and a scale radius, $r_s$, usually fixed to a large value for cluster scale DM haloes since SL cannot constrain it. Between $r = 0$ and $r = r_{\rm core}$, the mass density is constant. Then between $r = r_{\rm core}$ and $r = r_s$, the mass density is isothermal ($r^{-2}$), then it falls as $r^{-4}$ beyond $r_s$.

### 4.1. RATE & Nb

Two key parameters matter when it comes to the convergence of the LENSTOOL MCMC sampler (Jullo et al. 2007): the RATE parameter, associated to the burnin phase, and the number of iterations (Nb), associated to the sampling phase.

The burning phase is made of a sequential Monte Carlo process that gradually samples the priors to the posterior by the means of a tempered likelihood $\mathcal{L}_T = \mathcal{L}^{\frac{1}{T}}$ where $T$ is the likelihood temperature. $T$ is progressively decreasing from infinity to 1. Between each Monte Carlo process, $T^{-1}$ is updated based on the likelihood estimation at that temperature and the hyperparameter of the sampler RATE. The smaller the rate, the more the sampler will move slowly to the high-likelihood areas and will be less prone to miss a mode of the posterior. The larger Nb, the larger the number of iterations of the MCMC chains, the best the parameter space can be sampled. It is of good practice to lower the RATE and to increase Nb until the results obtained on the parameters are stable, in the sense that they remain consistent with each other, suggesting that the runs are not biased by the sampler's hyperparameters. The problem is that lowering the RATE and increasing Nb can become computationally very expensive, limiting our ability to thoroughly investigate the parameter space in an acceptable amount of time.

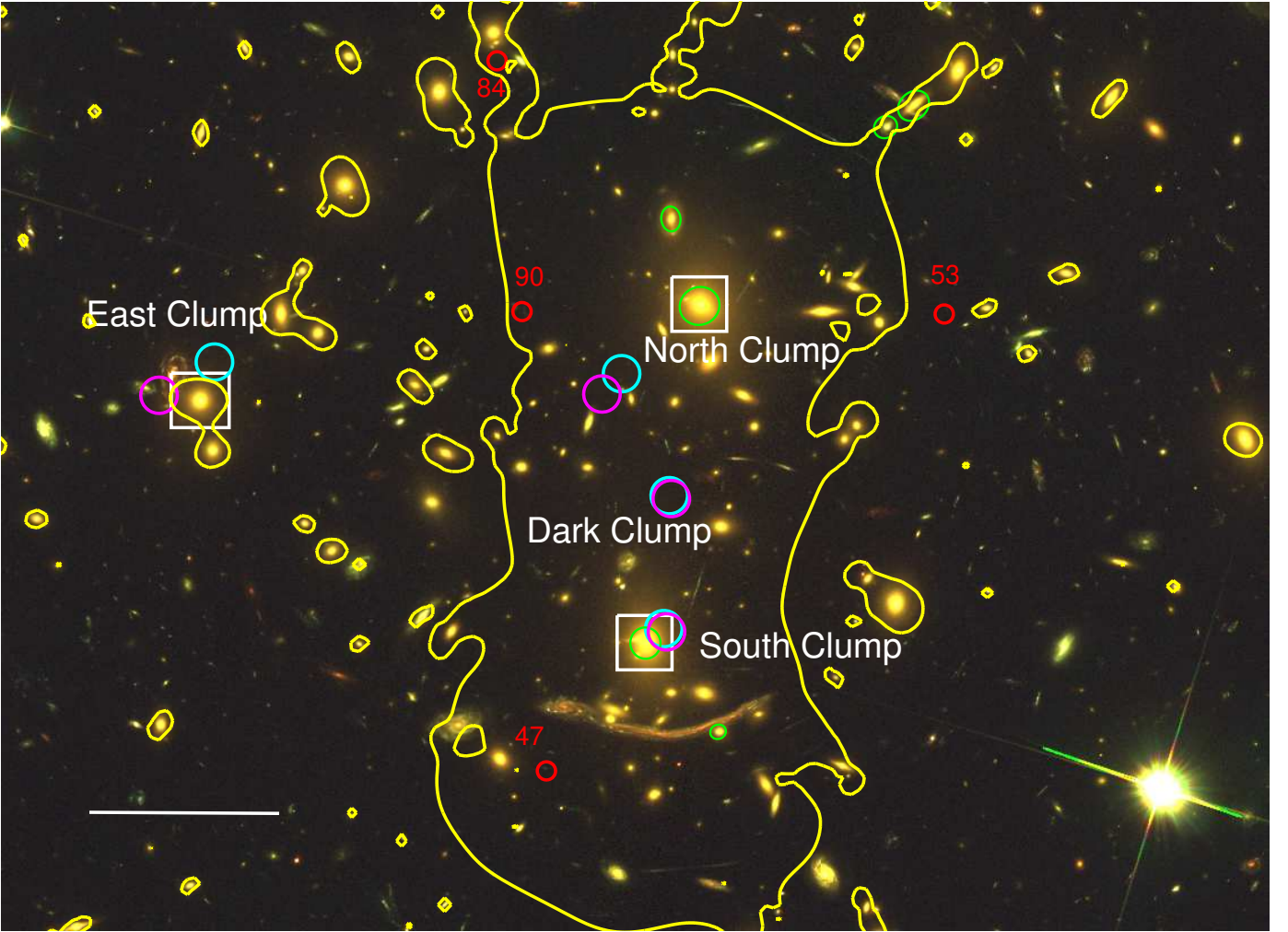

**Fig. 1.** Core of Abell 370 from BUFFALO data (F814/F606/F435 filters). Cyan circles correspond to the location of the four DM clumps from N23, and magenta circles those from model (ii) (Section 2.5). We draw in yellow the critical curve for $z = 2$ generated from model (ii). White squares show the location of each clump in the three DM clumps mass model (Section 5). Green ellipses show the cluster members which are optimised individually. We show in red the Lyman-$\alpha$ emitters from Claeyssens et al. (2022) for which we compute magnification (Section 7). North is up, east is left. The horizontal bar represents $20''$, which correponds to $\sim 100$ kpc at the redshift of Abell 370.

The N23 mass model was run with a RATE equal to 0.1 and Nb equal to 200, as were our own models presented in Section 2, in particular *model (ii)*. We therefore investigate *model (ii)* further, lowering the RATE and increasing Nb as long as we can afford it. In practice, we require a run to last at most three weeks on a dedicated modern 12-24 cores machine.

We show in Table 1 the values of RATE & Nb investigated, and the corresponding RMS. Note that we have two runs having the same RATE, Nb, and RMS ($0.7''$). However, the parameters differ, as shown in Fig. 2 and Fig. 3 where we present the corner plots for the parameters of each of the four mass clumps describing the mass distribution in Abell 370, for the different values of RATE & Nb investigated. Table 1 shows that the best fit model in term of RMS is not found using the values of RATE & Nb that allow for the most precise sampling, and that the RMS is ranging from $0.70''$ to $1.37''$. One would expect that the best RMS would be associated with the smaller RATE and the larger Nb, which is not the case here. We argue that this is due to the fact that our parametric description is not adapted to Abell 370. Besides, the parameter space is likely to have several minima and we may not be able numerically to properly sample it in order to converge to the global minimum. Fig. 2 and Fig. 3 show that the parameters of the DM clumps are very unstable from one run to another, being significantly different. Considering the variance between each run, we conclude that constraints are very loose for *all* parameters. Therefore, we do not present the parameters of the best fit model in a dedicated table as it is usually done in SL studies. We can split this class of 8 mass models between 4 having an RMS smaller or equal to 1″, and 4 having an RMS larger than 1″. However, we see no trends of the inferred values of the parameters of the mass clumps with respect to this dichotomy.

| RATE | Nb | RMS (″) |
|------|------|---------|
| 0.1  | 200  | 0.70 |
| 0.1  | 200  | 0.70 |
| 0.05 | 200  | 0.73 |
| 0.05 | 1000 | 1.04 |
| 0.05 | 1000 | 1.16 |
| 0.05 | 1000 | 1.28 |
| 0.05 | 1000 | 1.35 |
| 0.05 | 1000 | 1.37 |

**Table 1.** RMS obtained for the best-fit *model (ii)* given different values of RATE & Nb.

We also further investigate the reference model by N23 in the same way, lowering the RATE and increasing Nb to see how it influences the parameters of the DM clumps. Results are presented in Appendix B. We also present in Appendix A the results of such an exercise performed on galaxy cluster AS 1063, illustrating the behaviour of a run that has converged properly. In short, we find the results on the parameters describing AS 1063 to be very stable with lowering the RATE and increasing Nb, and that the corresponding RMS are equal.

### 4.2. The galaxy scale component

In the case of *model (i)*, where no spectroscopic prior is included, we derive a velocity dispersion of 153 km s$^{-1}$ (for a magnitude of 17.05), which is half of what is derived in *model (ii)* where a spectroscopic prior is considered ($\sigma = 307$ km s$^{-1}$, which is at the higher end of what is allowed by the spectroscopic prior). Regarding the individually optimised galaxies, in particular the BCGs, their properties are different from the one we would obtain following the spectroscopic prior. BCG-S and BCG-N have a similar magnitude (17.05 and 16.95 respectively), which would correspond to a velocity dispersion in the range 248±50 km s$^{-1}$ if we follow the spectroscopic prior. The optimised velocity dispersion are the following. For BCG-S, $\sigma = 190 \pm 23$ km s$^{-1}$ (*model(ii)*) and $\sigma = 144 \pm 31$ km s$^{-1}$ (*model(i)*). For BCG-N, $\sigma = 431 \pm 8$ km s$^{-1}$ (*model(ii)*) and $\sigma = 462 \pm 23$ km s$^{-1}$ (*model(i)*). Therefore, the two BCGs, which display a similar magnitude, end up having very different optimised velocity dispersions. BCGs in clusters constitute a particular class of galaxies which are not representative of the whole cluster member population and may not follow the same scaling relation. Whether or not these BCGs follow the spectroscopic prior, they should have a similar velocity dispersion given their similar magnitude. This shows that, when spectroscopically measured velocity dispersions become available for Abell 370, it will be relevant to consider them as additional constraints in the mass model.

### 4.3. DM & Light

The working hypothesis proposed in L22 is clearly not verified in the four clumps mass models discussed here (Fig. 1). Of these four DM clumps, a first one is associated with the bright cluster galaxy located in the South (BCG-S). The shape and orientation of the historical giant arc further South clearly indicates that BCG-S is dominating the total mass distribution in this area, fully justifying a DM clump sitting there. Then, a second DM clump is associated with the bright cluster galaxy located in the North (BCG-N). A bit further North, a long thin arc indicates that BCG-N dominates in this area, justifying the inclusion of a DM clump there. These two mass distributions are the dominant ones in Abell 370, as illustrated by a gravitational arc which is rather straight, located between BCG-S and BCG-N. A third DM clump is located in the Eastern part of the cluster, which we can associate with a bright galaxy located there (Fig. 1). Then a fourth DM clump is introduced in between BCG-S and BCG-N, with *no* luminous association. If the first and third DM haloes have positions which do coincide with the associated light distribution, the second one is located at $\sim$50 kpc from BCG-N, which is on the higher end of what is allowed by alternative DM models such as SIDM (see L22).

We then turn to the *total* projected mass, the quantity which is constrained by SL. We present in Fig. 4 the maps of the mean mass[1] obtained for the four mass models leading to an RMS smaller or equal to 1″ (Table 1). As expected, they are all in very good agreement. We note that the mass map associated to the larger RMS (1.04″) deviates from the other ones (having RMS equal to 0.73″ and 0.70″), which are pretty much indistinguishable from each other.

### 4.4. Conclusion on the four clumps mass model

We conclude that a four DM clumps mass model is able to reproduce accurately the observational constraints, with an RMS equal to 0.70″. As former models (L19, N23), its DM distribution does not follow the stellar component (Fig. 1, magenta circles). It features a "dark clump" and a significant offset between DM and light in the North. Besides, we show here that the parameters of each clump inferred from the different models reproducing the multiple images are poorly constrained. Still, the total projected mass is well constrained by the multiple images, but these multiple images are not able to provide much insight into the properties of DM in Abell 370, *e.g.* the number of DM clumps and their properties. As for the external shear that was reported in former studies, we consider that in such a complex and unconstrained parameter space, it is easy to accomodate for an extra component as an external shear.

---

[1] The mean mass map is generated from averaging the individual mass maps computed for each iteration of the MCMC sampler.

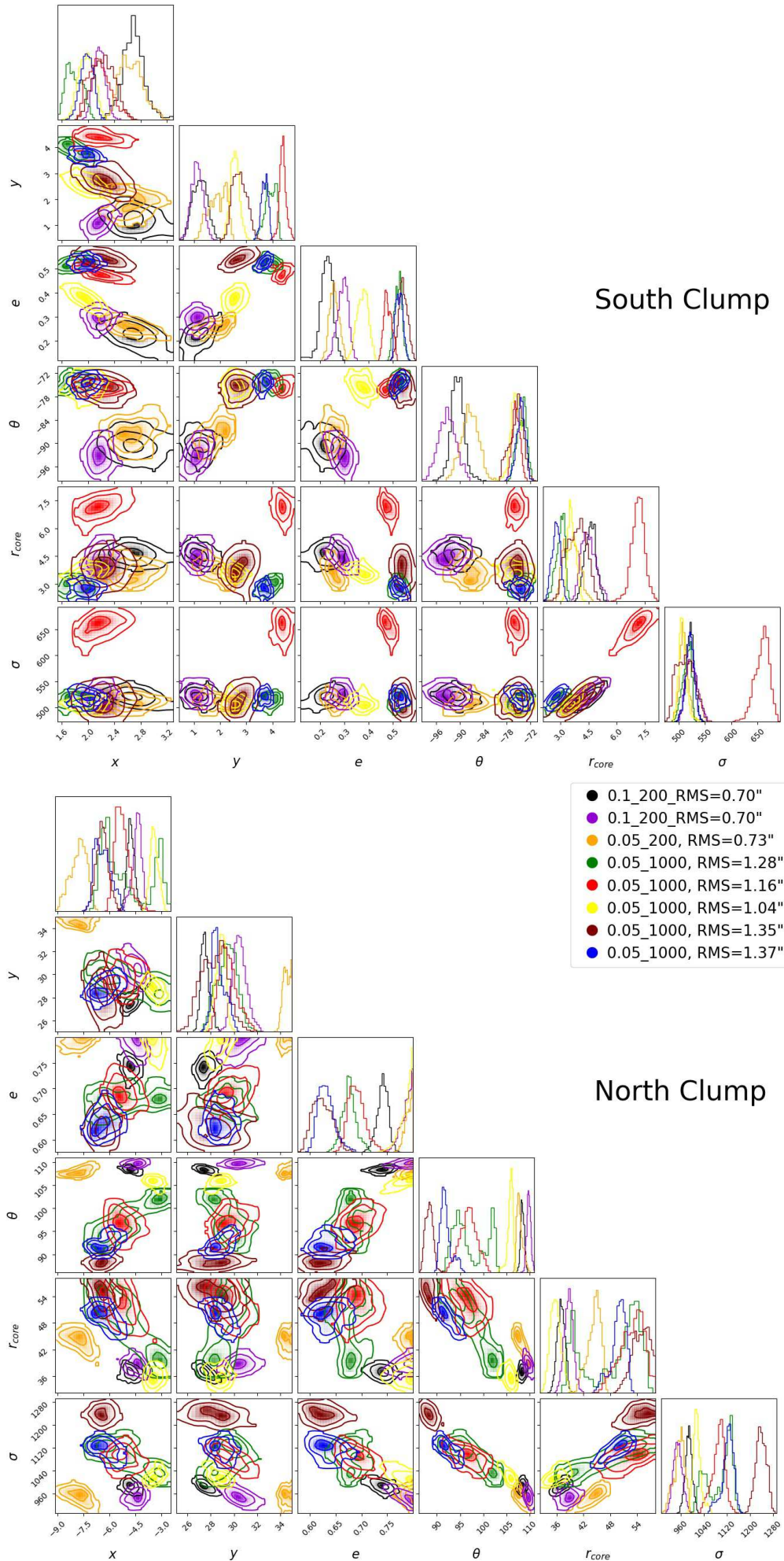

**Fig. 2.** Corner plots obtained for the parameters of the mass model, for the values of RATE & Nb indicated on the legend. *Top:* South clump; *bottom*: North clump. Parameters are significantly different.

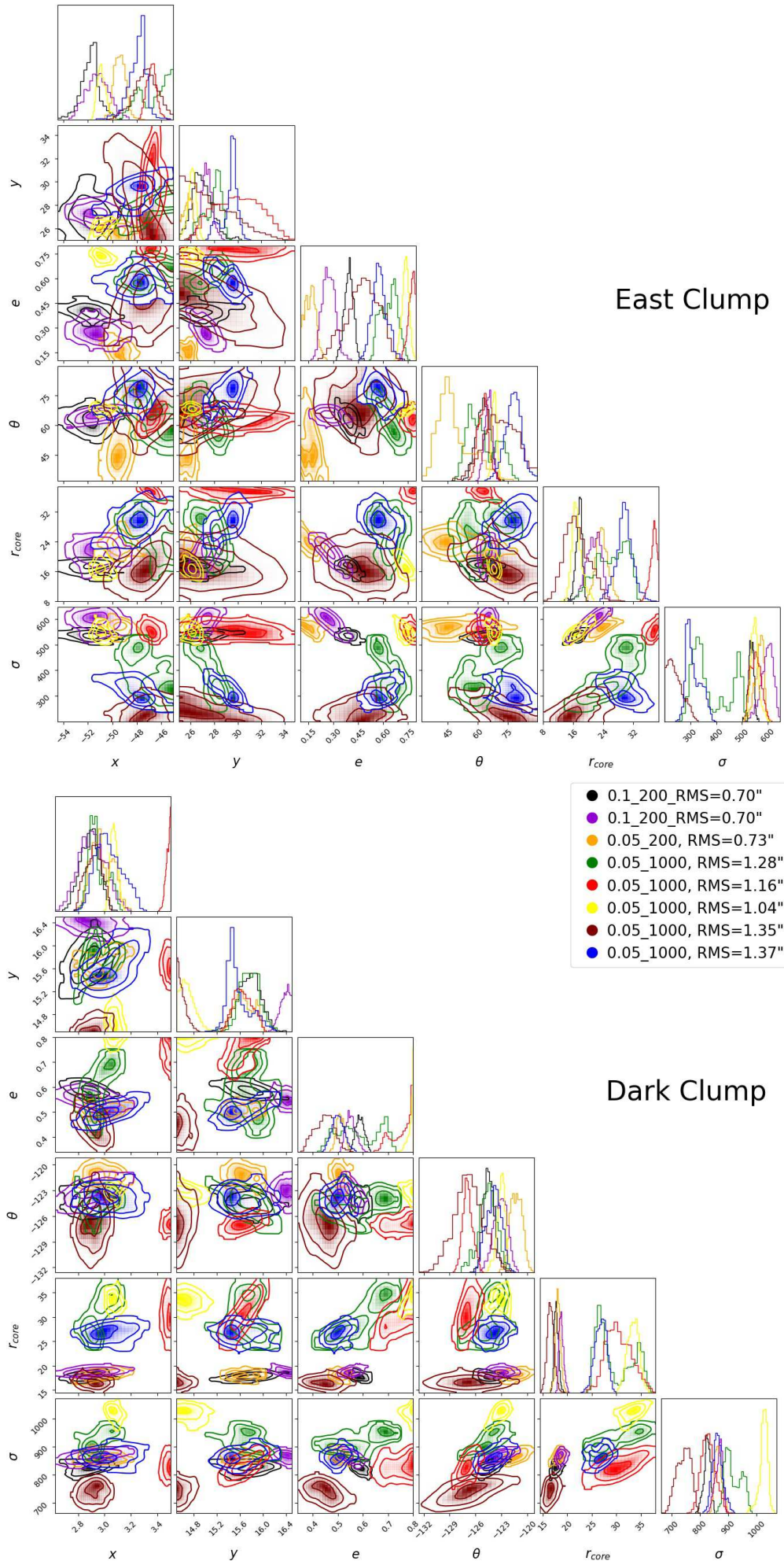

**Fig. 3.** Corner plots obtained for the parameters of the mass model, for the values of RATE & Nb indicated on the legend. *Top:* East clump; *bottom:* dark clump. Parameters are significantly different.

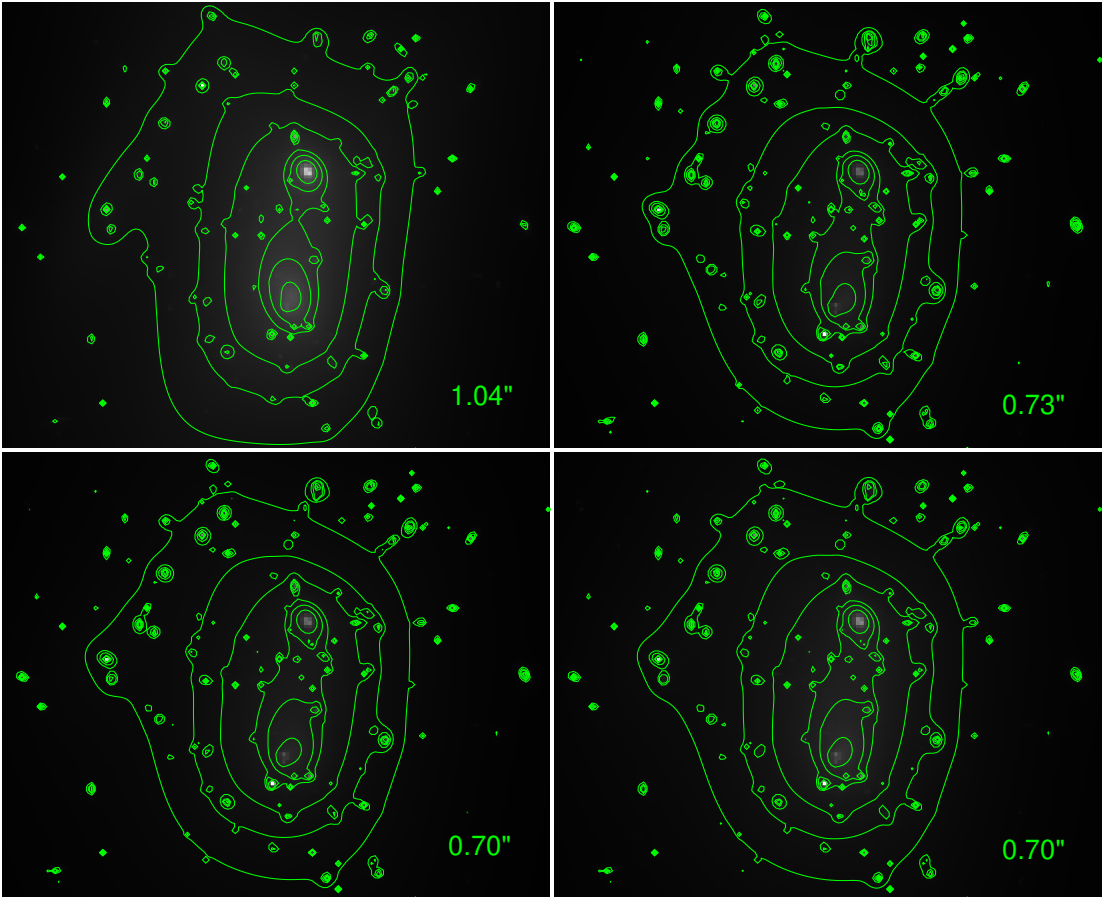

**Fig. 4.** Maps of the mean total projected mass derived from the models leading to an RMS smaller or equal to 1″, as reported on each map.

Actually, external shear components included in SL modelling is not always "shear" (Etherington et al. 2024), and different effects inherent to SL modelling can give rise to an effective external shear (see, *e.g.* Wagner 2019; Lin et al. 2022).

## 5. A Three DM clumps mass model

Following the working assumption proposed in L22, we now turn to investigate a mass model where each large scale DM clump is associated with a luminous counterpart. We propose that the total mass distribution in Abell 370 can be described by three DM clumps: one associated with BCG-S, one with BCG-N and one with the bright galaxy located in the east. The position of each clump is allowed to vary in a square of size 6×6 arcsec$^2$ centered on the corresponding bright galaxy (Fig. 1), typical of what is allowed by SIDM scenarios. On top of it, the galaxy scale perturbers are considered, using the spectroscopic priors discussed earlier. *No* external shear component is included in the modelling. As before, we investigate different values of RATE & Nb. These values, as well as the corresponding RMS, are shown in Table 2. We present in Fig. 5 the corner plots for the parameters of each of the three mass clumps describing the DM distribution in Abell 370, for the different values of RATE & Nb investigated. Note that since the positions of these mass clumps are forced to be coincident with their luminous counterpart, we do not show theses parameters in the figures, for clarity.

The best RMS found, equal to 2.24″, is larger compared to the four clumps mass model (0.70″). Similar to the behavior seen in the four clumps mass model, individual parameters in the three clumps model are very unstable from one run to another, and constraints are very loose. Moreover, an external shear component does not improve a three DM clumps mass model.

Substructures located outside the SL area can have an impact on the SL study, as discussed in Acebron et al. (2017). N23, from a weak lensing analysis of the BUFFALO data, studied the surroundings of the core of Abell 370 and actually detected several substructures. They provide both a "grid" and a parametric description of these substructures. We have been trying to take into account these substructures in our SL modelling, but this did not improve the situation, and the parameters of the main DM clumps remain loosely constrained. It is likely that adding more parameters in such an already complex and unconstrained

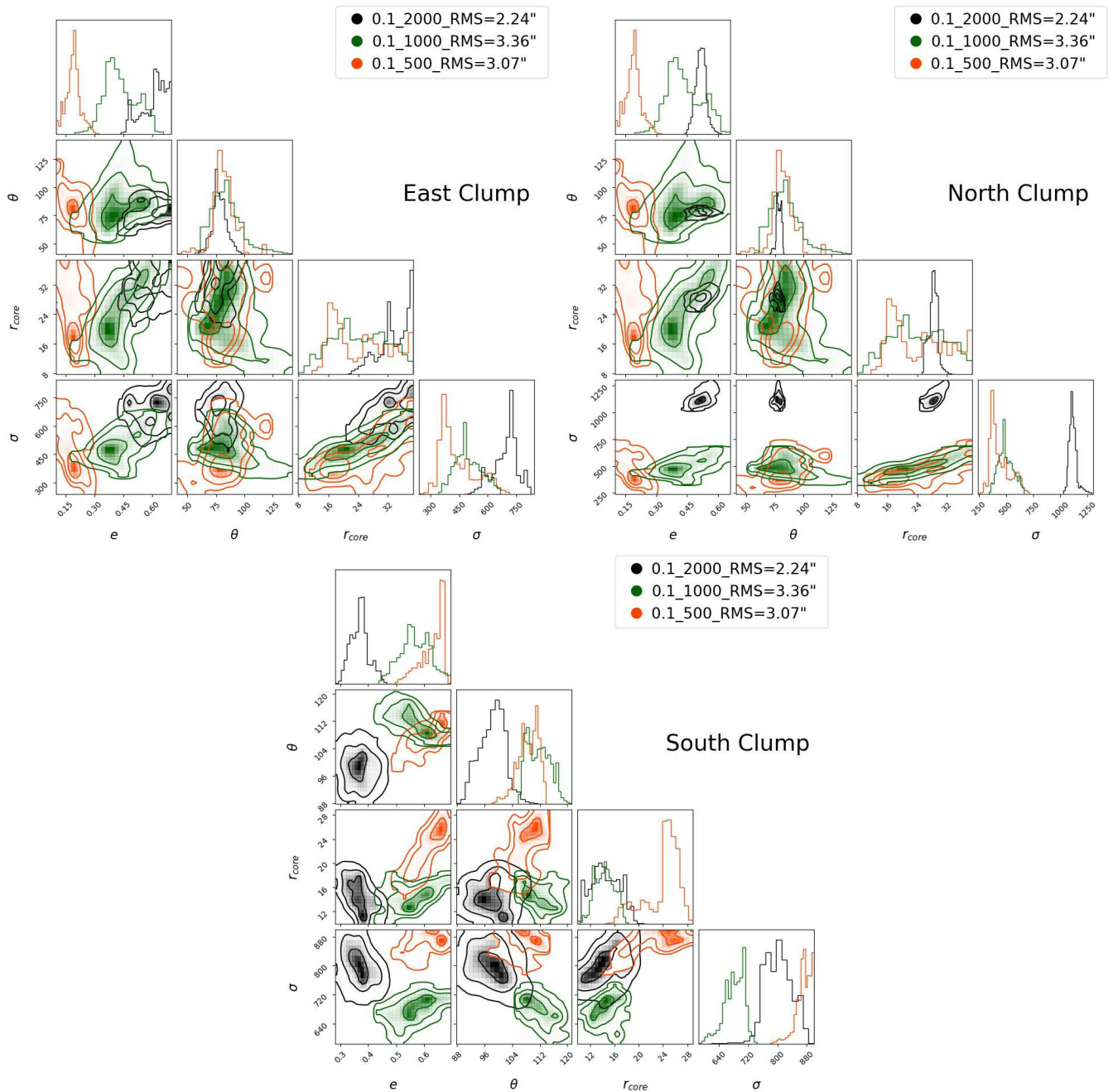

**Fig. 5.** Corner plots obtained for the parameters of each of the three mass clumps, for the values of RATE & Nb indicated on the legend. Parameters are significantly different.

parameter space cannot help. We reach the same conclusion in the four clumps mass model.

Following L22, we included a mild perturbation in the form of a B-spline surface in the lensing potential (Beauchesne et al. 2021) in order to see if this could improve the fit. Contrary to L22, we could not find a set of parameters for the B-spline surface that reduces the RMS. To ensure that it was not due to a numerical issue with respect to the complexity of A370 model, we tried to find such a surface while fixing the parametric model to the best-fit solution without more success.

To conclude, describing the mass distribution with a three DM clumps mass model, which positions are coincident with the light, leads to an RMS larger than 2″, three times the RMS obtained with a four DM clumps mass model. Moreover, the parameters of these three clumps are ill defined. The situation is therefore worse than the case where we would have such a large RMS but three well defined mass clumps. We therefore consider that we have failed to describe Abell 370 with a parametric model where the DM is traced by light.

We compare the maps of the mean *total* projected mass computed from this three DM clumps mass model with the map corresponding to our best four DM clumps mass model (Fig. 6). We find small deviations between these mass maps. In the same Figure, we also present the results correspond-ing to the N23 mass model, showing a very good agreement with our best fit mass model.

| RATE | Nb | RMS (″) |
|---|---|---|
| 0.1 | 500 | 3.07 |
| 0.1 | 1000 | 3.36 |
| 0.1 | 2000 | 2.24 |

**Table 2.** RMS obtained for the three clumps mass model, for different values of RATE & Nb.

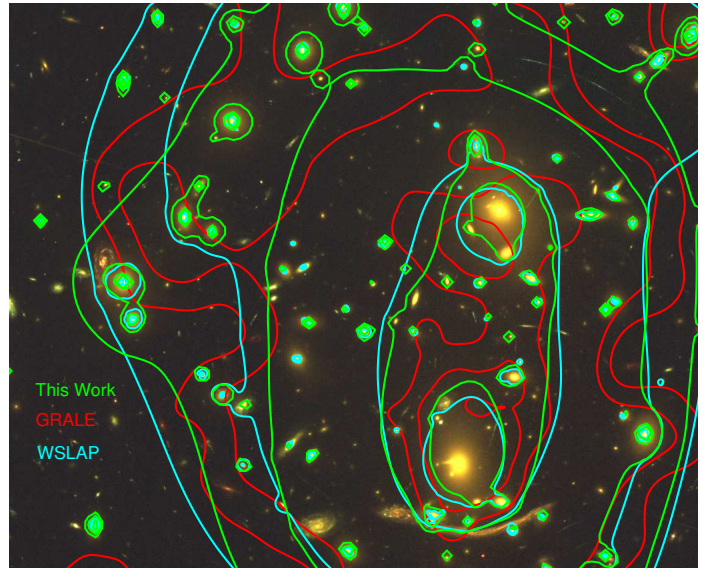

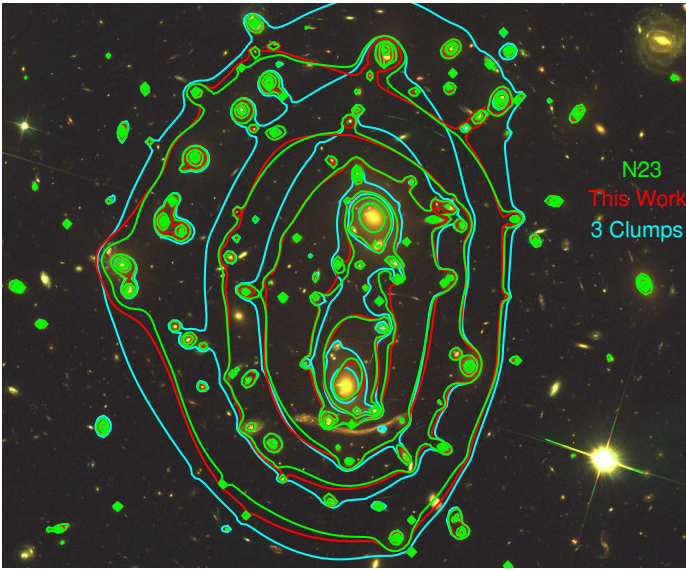

**Fig. 6.** Contours of the mean total projected mass derived from the following models: the three DM clumps mass model (cyan); the best four DM clumps mass model (red); N23 mass model (green). Contour values are the same for all three models.

**Fig. 7.** Convergence contours corresponding to the following models: the GRALE model (red), the best four DM clumps model (green), and the WSLAP+ model (cyan). Contour values are the same for all three models.

## 6. Discussion & Conclusion

### 6.1. Comparison with other studies

We compare our results with two other studies which have been using BUFFALO data: the one using WSLAP+ (Diego et al. 2018) and the one using GRALE (Ghosh et al. 2021). The authors shared their (total) convergence maps (which is proportional to the two dimensional surface mass density) on the dedicated BUFFALO website. We generate the same quantity from our best fit model and compare the three maps in Fig. 7.

In the South, the three reconstructions are in good agreement, with a peak coincident with BCG-S. In the North, our result is in good agreement with the WSLAP+ one, with a peak coincident with BCG-N, while the GRALE reconstruction features a peak slightly offset from BCG-N. The Eastern component is clearly sub-dominant in the total mass/light budget in all reconstructions. In this region, our reconstruction agrees well with the WSLAP+ one, while the GRALE reconstruction is slightly different.

Both WSLAP+ and GRALE methods are non parametric, but are using different modelling schemes that are described in detail in the corresponding publications. Here we just report one of the basic differences between the methods, to allow the reader to better understand the results of the comparison. In GRALE, no assumption is made about a relation between mass and light. In WSLAP+, the surface mass density is described by the combination of a soft component describing the DM and a compact component that accounts for the mass associated with individual cluster galaxies. Therefore, by construction, some mass is associated with cluster galaxies. If one considers the smooth DM component only, then the study by Diego et al. (2018) displays an offset between DM and the associated BCG. This offset disappears when considering the total mass, which is similar to what we find in this work.

We also compare our results with the one obtained by Cha & Jee (2023) on HFF data, using the non parametric algorithm MARS. This approach is closer to GRALE than to WSLAP+ since it does not use the galaxy distribution as prior. In their Fig. 4, they report the total mass contours inferred from their analysis. The reconstruction in the South is in good agreement with the other reconstructions discussed in this Section. In the North, their reconstruction displays a mass peak offset from BCG-N. This offset is slightly larger than the one found using GRALE.

### 6.2. Constraints on the DM distribution & interpretation of our results

A four DM clumps mass model is favoured compared to a three DM clumps mass model. The former DM distribution does not follow the stellar component whereas the latter does by construction. In both cases, the parameters of individual DM clumps are poorly constrained. We show that, in such a complex parameter space, degeneracies be-tween these parameters are large and little insight is gained on the properties of DM in Abell 370.

Still, the total projected mass is well constrained and is linked to the stellar component, the two main *total mass peaks* being coincident with the two BCGs (Fig. 6). We therefore conclude that *the total mass is traced by light in Abell 370.*

Having said that, we aim to discuss the underlying DM distribution, which, taken as such, might be misleading. What is the interpretation of this "dark clump": are we detecting a "dark clump"? What is the interpretation of the offset between the Northern DM clump and BCG-N, which is on the higher end of what is allowed by alternative DM models such as SIDM?

These interesting features (the "dark" clump and the offset) are clearly required by the data in order to reproduce the observed positions of the multiple images with a sub-arcsecond precision. We interpret these as not being "real" but rather being necessary to compensate the lack of realism of our parametric description of DM clumps during a cluster merging process. We, indeed, describe the DM component using idealised parametric mass profiles (*e.g.* dPIE or NFW). This description, though simple, can sometimes be reliable, which is remarkable. In the example of AS 1063 given in Appendix A (a unimodal cluster), the observed positions of the multiple images are well retrieved and the parameters of the DM component are well constrained. In Abell 370, such a simple description of the different DM components involved in the merging process might not fully capture the complex underlying physics, hence, some features, as the ones reported here, are needed to account for the deviations from our idealised parametric descriptions. In L22, we revisited three mass models featuring "misleading features" in former studies, and we were able to propose competitive mass models which do not require their inclusion. This is not the case for Abell 370. Despite this, we are able to remove the external shear component which was present in former LENSTOOL models. The last parametric study by Li et al. (2024) goes in this direction, in the sense that their DM description contains "misleading features" as listed in Section 2, needed to reproduce accurately the multiple images. Actually, some former works on Abell 370 report such features, as summarized in Table 2 by Ghosh et al. (2021).

Even in the JWST era, where hundreds of multiple images are observed, SL mass reconstructions still suffer from degeneracies (Lasko et al. 2023; Liesenborgs et al. 2024), in particular in merging clusters, and caution and criticism should be taken when reading and interpreting the results of any SL model. Furthermore, authors could discuss more the limitations of their models, and help the reader to understand and interpret their results. We therefore encourage caution and criticism on the outputs of parametric SL modelling. This is our main conclusion. We finish this paper by discussing the implications of our results for using Abell 370 as a gravitational telescope.

# 7. Implication for high $z$ studies

The findings of Section 4 might have implications for using Abell 370 as a gravitational telescope, since the different models reproducing the SL constraints are likely to provide different magnification estimates for background lensed objects, as well as different critical curves.

| ID | RA | Dec | $z$ |
|---|---|---|---|
| 53 | 39.962233 | -1.57206 | 4.916 |
| 47 | 39.974424 | -1.58609 | 3.130 |
| 90 | 39.975065 | -1.57212 | 3.159 |
| 84 | 39.975825 | -1.56444 | 6.173 |

**Table 3.** Coordinates and redshift of the four Ly$\alpha$ emitters for which we compute magnification values.

## 7.1. Magnification estimates

Quantifying this effect on global properties such as the luminosity function derived for background objects is clearly beyond the scope of this paper and might be adressed in a forthcoming publication, but to illustrate our claim, we compute the magnification derived on a few background galaxies. We consider four objects selected from the sample of 100 Lyman-$\alpha$ emitters lensed by Abell 370 reported by Claeyssens et al. (2022). Their coordinates and redshifts are given in Table 3, and their locations are shown in Fig. 1. This selection is rather arbitrary; these objects do not have any particuliar characteristics within the sample. We compute the magnification experienced by these objects using the different mass models investigated in this paper presenting an RMS smaller or equal to $1.0''$, *i.e.* the ones that reproduce the SL constraints with the greatest accuracy. This corresponds to the six models constituting the N23 like class of models (Appendix B), and four models out of the eight models presented in Section 3. This makes a total of ten mass models. Results are shown in Fig. 8. We see that the magnification for a given object depends on the mass model used, and that if one was to consider a class of models reproducing the SL constraints instead of one single model, uncertainties on the magnification estimate would increase. We see that for object 53, we only have 5 curves instead of 10. In fact, for the 5 mass models not shown here, the magnification is unconstrained, having error bars larger than a few hundreds, hence not shown for clarity. This is typical of the kind of objects that one would not use to study the high redshift Universe, since the computed magnification values are very unstable. We see here that this criterion depends on the mass model considered: depending on the mass model used, one would accept or reject such an object. Uncertainties on the magnification are much larger (sometimes unconstrained) when using the three DM clumps mass model, and not shown.

## 7.2. Critical curves: the "Dragon arc"

The historical giant arc at $z = 0.725$, dubbed the "Dragon arc" has been the subject of recent interest since JWST observations have led to the discovery of $\sim 50$ microlensed stars in this strongly-lensed galaxy (Kelly et al. 2022; Fudamoto et al. 2024). Considering the class of models leading to an RMS smaller or equal to $1''$, we compute the critical curves at $z = 0.725$ and present them on Fig. 9, together with the location of the microlensing events and the critical curves obtained using WSLAP+. This illustrates how the critical curves differ from one model to another one. The distribution of microlensing events depends mostly on i) the position of the critical curve, ii) the surface mass density of microlenses, iii) the presence of millilenses (for instance unresolved globular clusters in the lens plane), and iv) the bright-end slope of the stellar luminosity function of the background lensed galaxy (see for instance Diego et al. 2024). Future observations of new transients in this arc will significantly increase the statistics of microlensing events allowing to reverse engineer the distribution of microlensing events and narrow down the precise location of the macromodel critical curve. This in turn can be used to improve the precision of the lens models in this portion of the cluster.

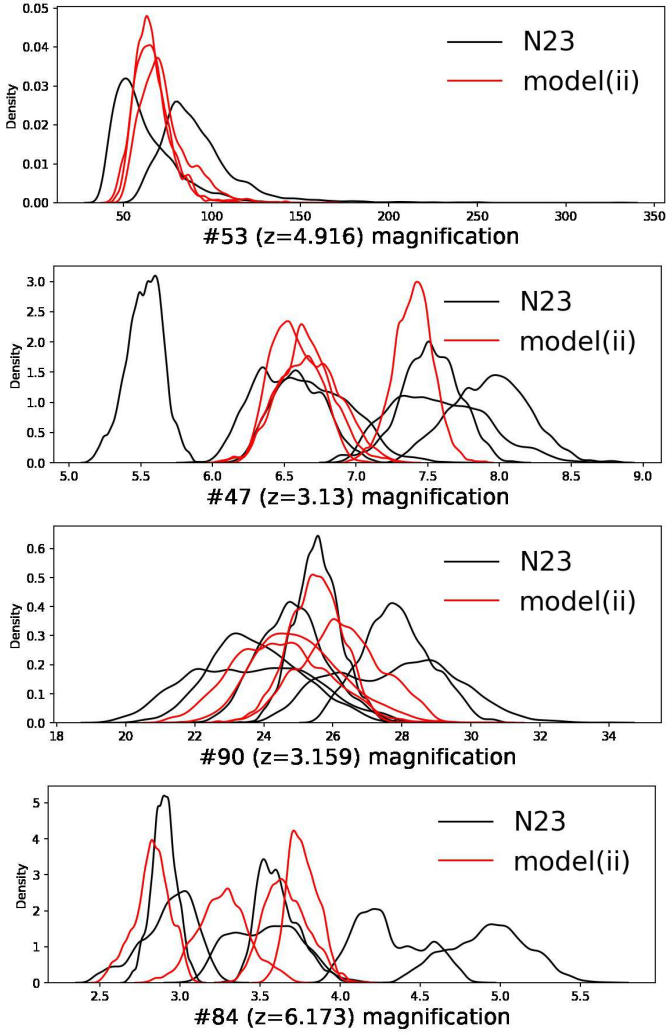

**Fig. 8.** Magnification computed for the four background Lyα emitters using all models presenting an RMS smaller or equal to $1''$.

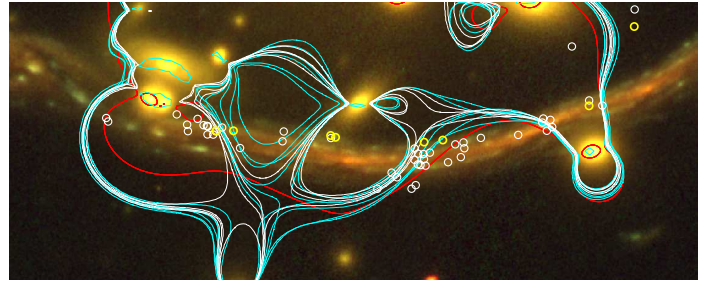

**Fig. 9.** Critical curves at $z = 0.725$ computed for the class of models leading to an RMS smaller or equal to $1''$, both from the N23 class of models (cyan) and from model(ii) in white. In red, critical curves derived with WSLAP+. Note that all these models are using the N23 constraints. We show the location of the microlensing events reported by Fudamoto et al. (2024) as white circles and by Kelly et al. (2022) in yellow.

### 7.3. Clusters as gravitational telescopes

Related to this issue is the uncertainty on the derived magnification maps in multimodal clusters as Abell 370 or MACS 0717 (for which similar conclusions where drawn, see Limousin et al. 2016), and the implications for high redshift studies in cluster fields. One is tempted to select complex disturbed clusters since they provide a larger area of high magnification than "simple" relaxed clusters. This has been the strategy of the Hubble Frontier Fields initiative, where 5 of the 6 targets are complex clusters. The problem associated with this approach is that we end up with modelling uncertainties that translates into large uncertainties on the magnification estimates and that the area where the magnification can be considered as reliable drops significantly (see the discussion by Limousin et al. (2016) in the case of MACS 0717). Therefore, dedicated programs targetting "simple" clusters might also be considered to build an interesting strategy. The answer cannot be definite and both populations (complex multimodal and simple regular clusters) are worth observing, as long as we are aware of pros and cons, in particular, not largely under-estimating the errors on the magnifications in the case of complex clusters.

### Acknowledgement

ML acknowledges the Centre National de la Recherche Scientifique (CNRS) and the Centre National des Etudes Spatiale (CNES) for support. This work was performed using facilities offered by CeSAM (Centre de donnéeS Astrophysique de Marseille). J.M.D. acknowledges the support of project PID2022-138896NB-C51 (MCIU/AEI/MINECO/FEDER, UE) Ministerio de Ciencia, Investigación y Universidades. MJ is supported by the United Kingdom Research and Innovation (UKRI) Future Leaders Fellowship 'Using Cosmic Beasts to uncover the Nature of Dark Matter' (grant numbers MR/S017216/1 and MR/X006069/1). AA acknowledges financial support through the Beatriz Galindo program and the project PID2022-138896NB-C51 (MCIU/AEI/MINECO/FEDER, UE), Ministerio de Ciencia, Investigación y Universidades. DJL is partially supported by STFC grants ST/T000244/1 and ST/W002612/1. He is also partially supported by the United Kingdom Research and Innovation (UKRI) Future Leaders Fellowship 'Using Cosmic Beasts to uncover the Nature of Dark Matter' (grant number MR/S017216/1). LF acknowledges support by grant No. 2020750 from the United States-Israel Binational Science Foundation (BSF) and grant No. 2109066 from the United States National Science Foundation (NSF), by the Israel Science Foundation Grant No. 864/23, and by the Ministry of Science & Technology, Israel. P.N. acknowledges support from DOE grant #DE-SC0017660. This research was supported by the International Space Science Institute (ISSI) in Bern, through ISSI International Team Project #476 (Cluster Physics From Space To Reveal Dark Matter).

## Appendix A: AS 1063: Example of a run that has converged.

We consider the mass model published in L22, imposing the working hypothesis proposed in this paper (positions of the DM clumps coinciding with the light and ellipticity lower than 0.7) and vary RATE & Nb. Table A.1 shows the results in term of RMS for the different values investigated, and we show in Fig. A.1 the corner plots inferred for each parameter of the mass model, obtained for each value of RATE & Nb. These parameters are the following: position, ellipticity, position angle, core radius and velocity dispersion of the main dominant large scale DM clump associated with the BCG. Then, we have the scale radius and velocity dispersion of the galaxy scale perturbers ($r_{pot}$ and $\sigma_{pot}$ respectively). We see that the results are very stable, and do not depend on the RATE & Nb: RMS for all runs are similar, and PDFs for all parameters are consistent with each other.

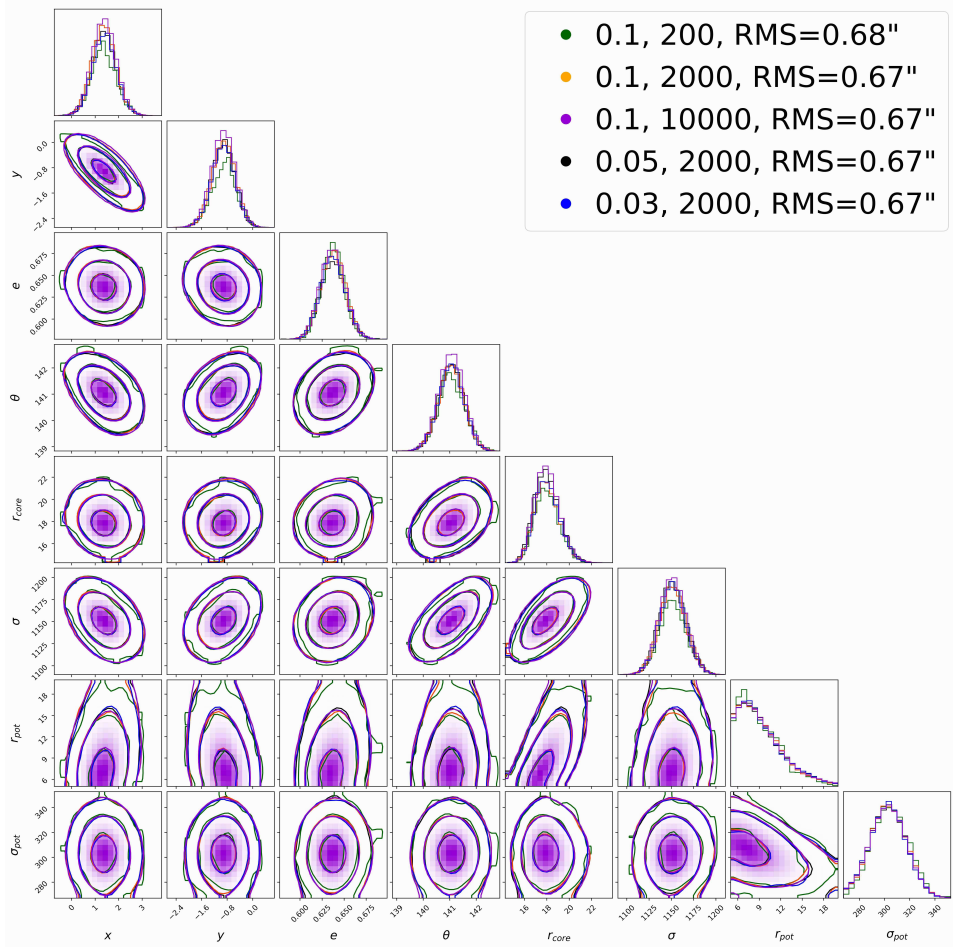

**Fig. A.1.** Corner plots obtained for the parameters of the mass model, for the values of RATE & Nb indicated on the legend.

| RATE | Nb | RMS ($''$) |
|------|-------|------|
| 0.1  | 200   | 0.68 |
| 0.1  | 2000  | 0.67 |
| 0.1  | 10000 | 0.67 |
| 0.05 | 2000  | 0.67 |
| 0.03 | 2000  | 0.67 |

**Table A.1.** RMS obtained on AS 1063 for different values of RATE & Nb.

clumps are very unstable from one run to another, similar to what is found in the investigation of *model (ii)* exposed in Section 3.1.

We then come to the comparison of the *total* projected mass distributions derived from these models, presenting in Fig. B.3 the results for the N23 and the N23 like class of models, showing a very good agreement between the mass maps. This agreement is expected, given that SL is sensitive to the total projected mass.

## Appendix B: Investigating the N23 model further

We consider the mass model proposed by N23, with RATE=0.1 & Nb=200, RMS=0.90$''$. We then vary RATE & Nb as summarized in Table B.1, where we report the RMS of each run: this constitutes the N23 like class of models. We show in Fig. B.1 and Fig. B.2 the corner plots obtained for each of the four DM clumps parameters.

The best model in terms of RMS is not found for the best values of RATE & Nb, but the RMS values are stable. The corner plots show that the parameters of the DM

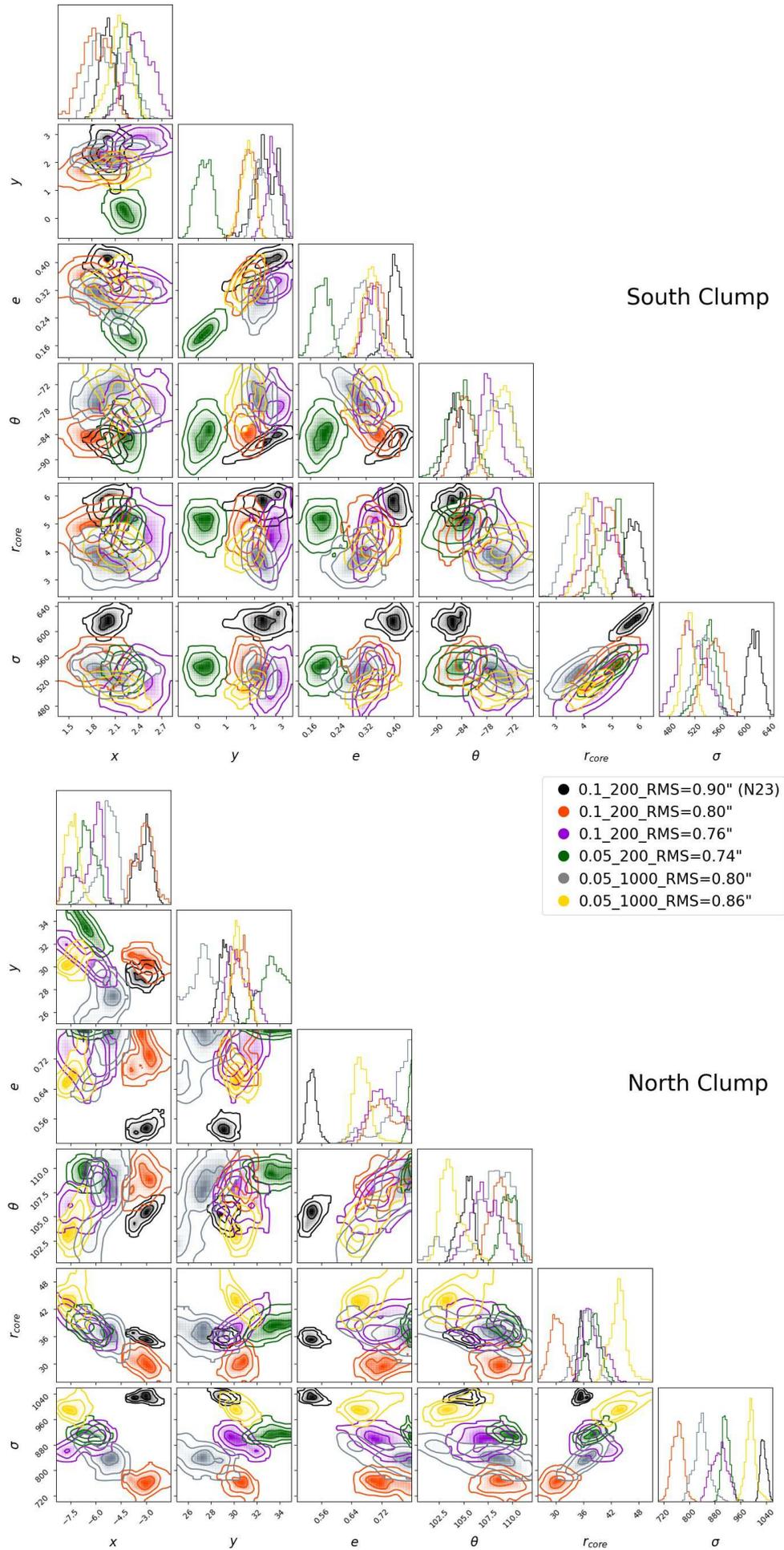

**Fig. B.1.** Corner plots obtained for the parameters of the mass model, for the values of RATE & Nb indicated on the legend. *Top:* South clump; *bottom:* North clump.

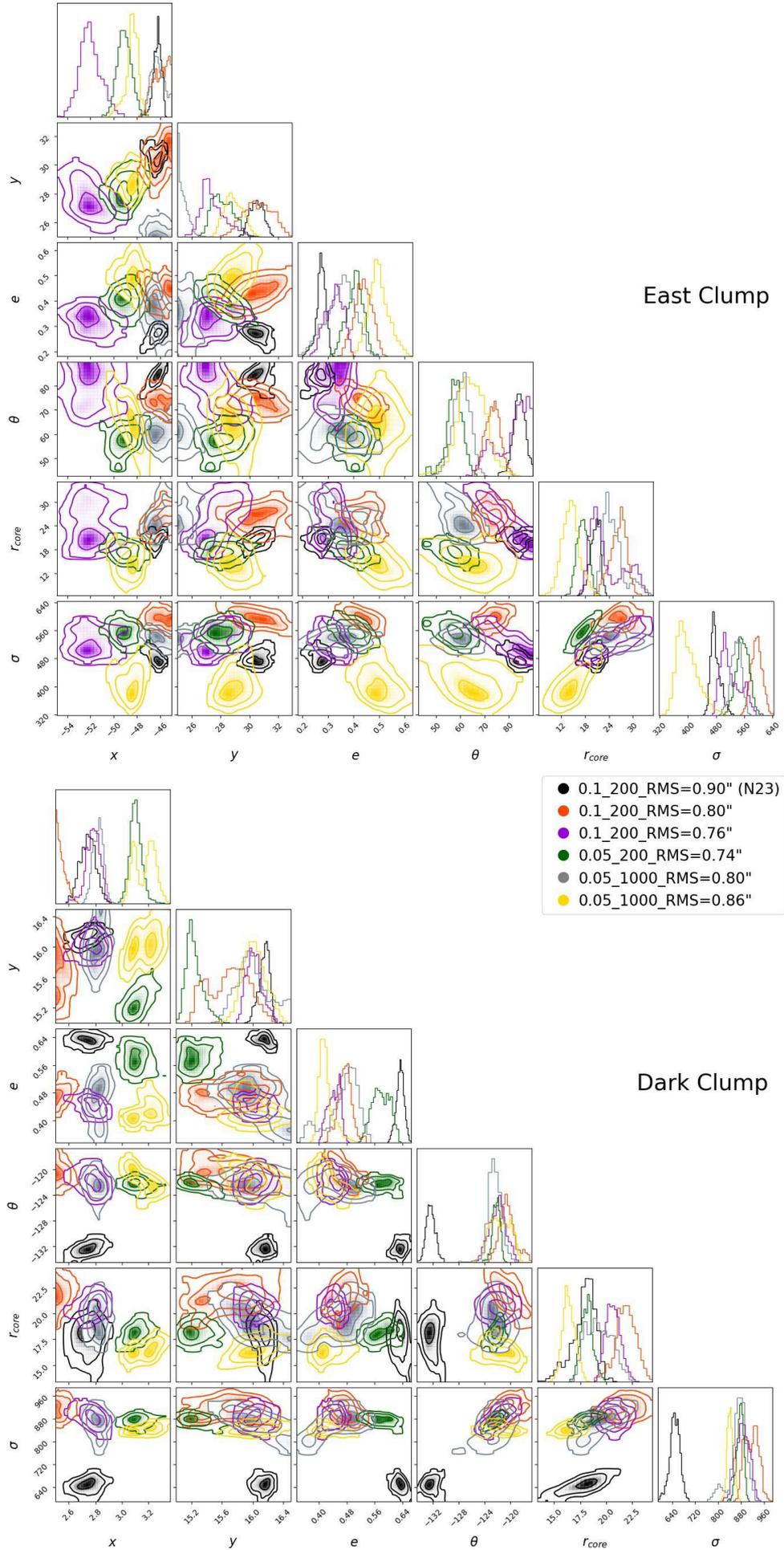

**Fig. B.2.** Corner plots obtained for the parameters of the mass model, for the values of RATE & Nb indicated on the legend. *Top:* East clump; *bottom:* dark clump.

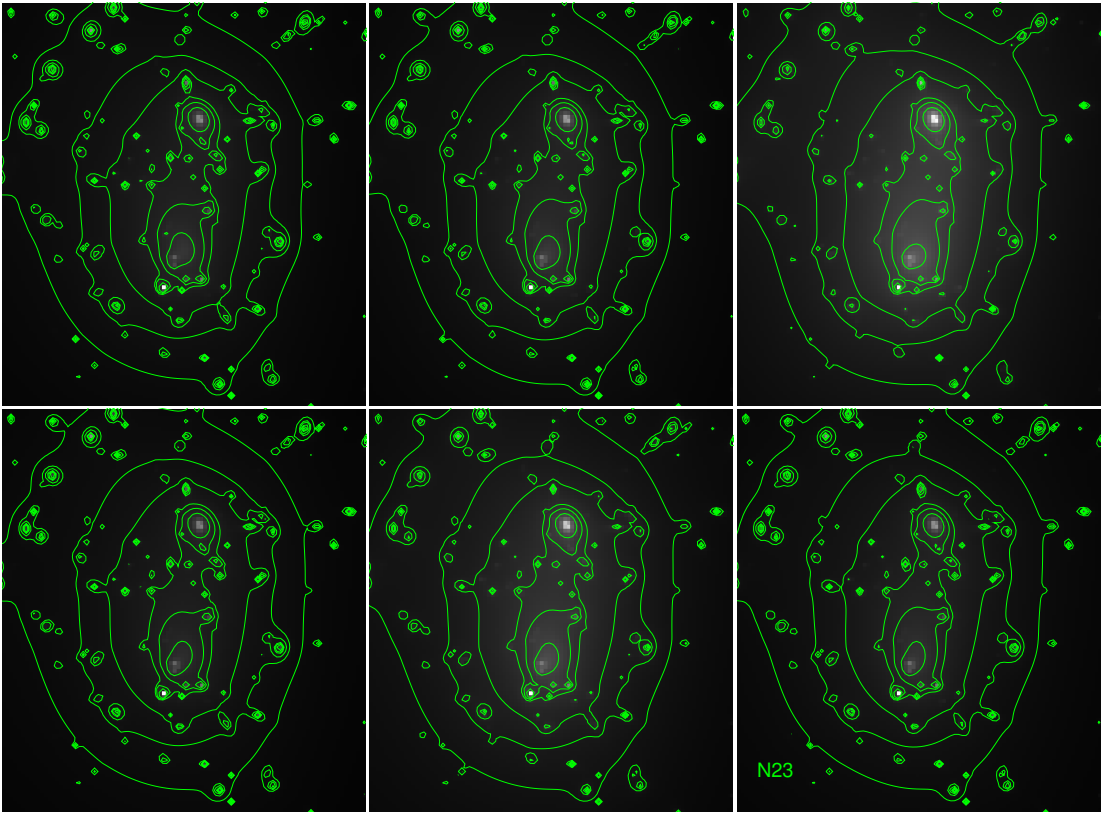

**Fig. B.3.** Maps of the total projected mass derived from the N23 (lower right), and the N23 like class of models.

| RATE | Nb | RMS ($''$) |
|---|---|---|
| 0.1 | 200 | 0.90 (N 23) |
| 0.1 | 200 | 0.80 |
| 0.1 | 200 | 0.76 |
| 0.05 | 200 | 0.74 |
| 0.05 | 1000 | 0.80 |
| 0.05 | 1000 | 0.86 |

**Table B.1.** RMS obtained for N23 like models, for different values of RATE & Nb.